\documentclass[iop,revtex4]{emulateapj}
\usepackage{color}
\usepackage[breaklinks,colorlinks,citecolor=blue,linkcolor=magenta]{hyperref}

\shorttitle{Star-grazing Comets in the 49~Ceti Disk}
\shortauthors{Miles et al.}

\newcommand{\cet}{49~Cet}
\newcommand{\bp}{$\beta$~Pic}

\newcommand{\kms}{km~s$^{-1}$}

\begin{document}

\title{UV Spectroscopy of Star-grazing Comets Within the 49~Ceti Debris Disk}

\author{Brittany E. Miles}
\affil{Dept.\ of Physics and Astronomy, University of California, Los Angeles, 
430 Portola Plaza, Box 951547, Los Angeles, CA 90095, USA}

\author{Aki Roberge}
\affil{Exoplanets \& Stellar Astrophysics Lab, NASA Goddard Space Flight Center, Code 667, Greenbelt, MD 20771, USA \email{Aki.Roberge@nasa.gov}}

\and

\author{Barry Welsh}
\affil{Eureka Scientific, 2452 Delmer, Suite 100, Oakland, CA 96002, USA}

\begin{abstract}

We present analysis of time-variable, Doppler-shifted absorption features in far-UV spectra of the unusual 49~Ceti debris disk.
This nearly edge-on disk is one of the brightest known, and is one of the very few containing detectable amounts of circumstellar gas as well as dust. 
In our two visits of \emph{Hubble Space Telescope} STIS spectra, variable absorption features are seen on the wings of lines arising from \ion{C}{2} and \ion{C}{4}, but not for any of the other circumstellar absorption lines.
Similar variable features have long been seen in spectra of the well-studied $\beta$~Pictoris debris disk and attributed to the transits of star-grazing comets.
We calculated the velocity ranges and apparent column densities of the 49~Cet variable gas, which appears to have been moving at velocities of tens to hundreds of km~s$^{-1}$ relative to the central star.
The velocities in the redshifted variable event seen in the second visit show that the maximum distances of the infalling gas at the time of transit were about 0.05 to 0.2~AU from the central star.
A preliminary attempt at a composition analysis of the redshifted event suggests that the C/O ratio in the infalling gas is super-solar, as it is in the bulk of the stable disk gas.
%\url{https://docs.google.com/a/g.ucla.edu/document/d/1N7hMMGyrwGcn6kS5cqB1lDDpvqy96QEIkH0hJiNVti8/edit}

\end{abstract}


%%%%%%%%%%%%%%%%%%%%%%%%%%%%%%%%%%%%%%%%%%%%%%%%%%%%%%%%%%%%%%%%%%%%%%%%%%%%%%%%%%%%%%%%%%

\section{Introduction}

A debris disk is a type of circumstellar (CS) disk that is primarily composed of dust grains coming from planetesimals analogous to comets and asteroids in the Solar System (the Solar System in fact hosts a debris disk, referred to as the zodiacal dust). In contrast to younger protoplanetary disks, debris disks contain modest amounts of dust and are optically thin. They are typically discovered via infrared (IR) photometry of excess thermal emission, as the CS dust absorbs short-wavelength light from the central stars and reprocesses it to longer wavelengths. 
 
\object[HD9672]{49~Ceti} is a nearby (61~pc) young A1V star that hosts a bright debris disk containing an unusually large amount of carbon monoxide gas \citep{Dent:2005,Hughes:2008}. The presence of both gas and dust grains in the disk led to the idea of 49~Cet being a late-stage transitional disk, just on the verge of becoming a typical gas-depleted debris disk \citep{Hughes:2008}. Further work constrained the age of 49~Cet by associating it with the Argus young moving group \citep{Zuckerman:2012}. This indicates that the age of 49~Cet is $\sim 40$~Myr, making it unlikely that the gas in the disk is a remnant left over from stellar formation, but rather is constantly replenished. \citet{Zuckerman:2012} attributed the gas to frequent collisions of icy comet-like bodies in the disk.

More information for the comet collision scenario has come from far-IR spectra of 49~Cet obtained with the \emph{Herschel Space Observatory} \citep{Roberge:2013}; in contrast to far-IR observations of protoplanetary disks, \ion{C}{2} emission was detected but \ion{O}{1} emission was not. More recently, far-ultraviolet (far-UV) spectra of 49~Cet showed many strong atomic absorption lines arising from circumstellar gas \citep{Roberge:2014}. Analysis of those lines indicated a super-solar abundance of carbon to oxygen in the main disk gas, which would be highly unusual for protoplanetary gas around a nearby main sequence star like 49~Cet.
%No line of sight CO absorption was detected in the far-UV spectra, despite previous %observations of strong sub-mm CO emission. This discrepancy in CO detection implied that %the disk is not exactly edge-on to our line of sight. 
The far-UV spectra of 49~Cet presented in \citet{Roberge:2014} are also analyzed in this current paper.
 
If a large population of comets exists within a disk, then some of those bodies may get perturbed towards the host star and sublimate when they get close enough. 
For a disk like 49~Cet, which is nearly edge-on from our vantage point \citep{Hughes:2008}, some proportion of the comets may pass through our line of sight to the central star.
During these transits, they would produce Doppler-shifted absorption lines superimposed on the stellar spectrum.
Such spectral features, which are variable on the timescale of hours to days, are a long-known and well-studied aspect of the famous $\beta$~Pictoris debris disk \citep[e.g.][]{Beust:1990,Kiefer:2014}.
 
Several other stars have shown Doppler-shifted, variable absorption features in optical spectra of the \ion{Ca}{2} K line, including 49~Cet \citep{Montgomery:2012}. 
However, with the exception of $\beta$~Pic, no other debris disk has had multiple UV spectroscopic observations of the infalling gas done over short time scales. In this paper, we present analysis of infalling gas events seen in the far-UV spectra of 49~Cet, and strengthen their link to extrasolar comets by examining the gas dynamics and composition. 

%%%%%%%%%%%%%%%%%%%%%%%%%%%%%%%%%%%%%%%%%%%%%%%%%%%%%%%%%%%%%%%%%%%%%%%%%%%%%%%%%%%%%%%%%%

\section{Data}

High-resolution ($R = \lambda / \Delta \lambda = 228,000$) far-UV spectra of 49~Cet were obtained with the \emph{Hubble Space Telescope} (\emph{HST}) Space Telescope Imaging Spectrograph (STIS) on 2013-08-11 and 2013-08-16. 
A full description of the dataset appears in \citet{Roberge:2014}.
The radial velocity of the central star is $\sim 12.2$~\kms\ \citep{Hughes:2008}.
To increase the S$/$N of the data, each spectrum was rebinned by a factor of 3.
The spectrum from the second visit was interpolated onto the wavelength scale of the first visit. 

\begin{figure}
\plotone{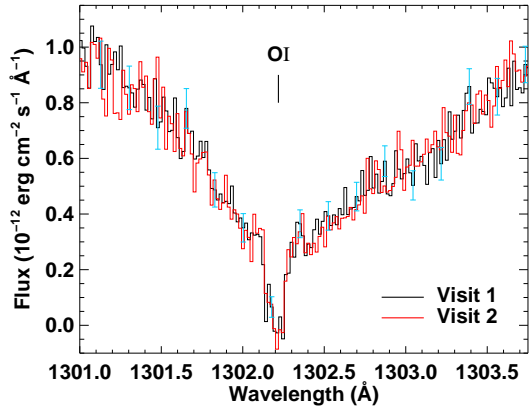}
\caption{An unvarying \ion{O}{1} line showing strong circumstellar absorption in the \cet\ STIS data. The Visit~1 spectrum is plotted with a black solid line, while the Visit~2 spectrum is plotted with the red solid line.
\label{fig:unvarying}}
\end{figure}

\begin{figure}
\plotone{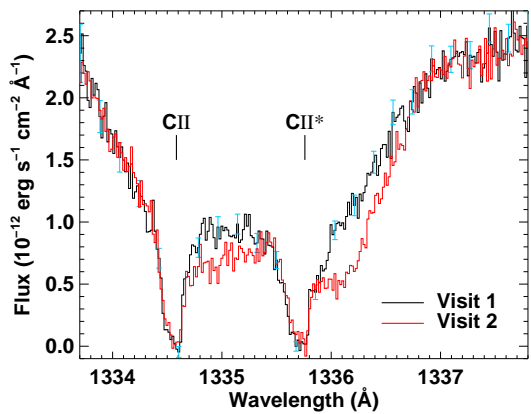}
\caption{The variable \ion{C}{2} lines in the \cet\ STIS data. The Visit~1 spectrum is plotted with a black solid line, while the Visit~2 spectrum is plotted with a red solid line. Excess redshifted absorption is clearly visible in the Visit~2 spectrum.
\label{fig:cii}}
\end{figure}

\begin{figure}
\plotone{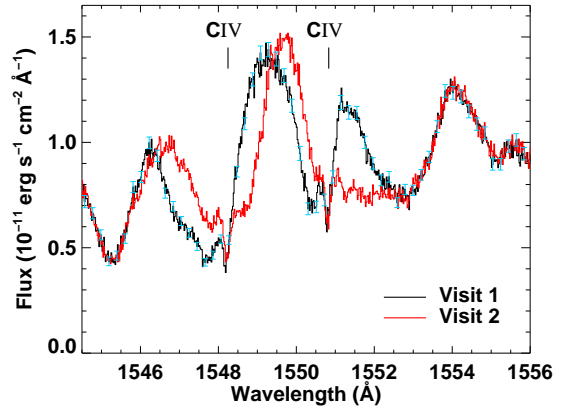}
\caption{The variable \ion{C}{4} lines in the \cet\ STIS data. The Visit~1 spectrum is plotted with a black solid line, while the Visit~2 spectrum is plotted with the red solid line.
As for the \ion{C}{2} lines, excess redshifted absorption arising from infalling gas is visible in the Visit~2 spectrum.
However, in this case, there is also excess blueshifted absorption in the Visit~1 spectrum, showing that two very different star-grazing comet events were recorded. \label{fig:civ}}
\end{figure}

\begin{figure}
\plotone{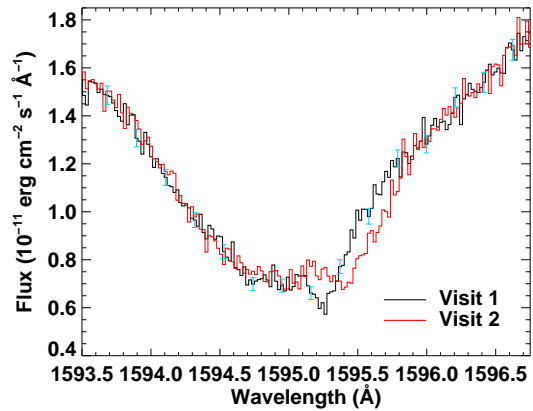}
\caption{Unidentified variable absorption feature. \label{fig:mystery}}
\end{figure}

Most of the absorption features in the spectra did not vary between the two visits, showing only stable absorption associated with the main \cet\ gas disk. 
A plot of an unvarying \ion{O}{1} absorption line appears in Figure~\ref{fig:unvarying}. 
However, the wings of features arising from \ion{C}{2}, \ion{C}{2}* (Figure~\ref{fig:cii}), and \ion{C}{4} (Figure~\ref{fig:civ}) did vary significantly between the two visits. 
Furthermore, there is one unidentified variable feature in the wavelength range 1594.0 -- 1596.5~\AA\ (Figure~\ref{fig:mystery}).

% Comment from Aki: This info below is good to include in a undergrad 
% thesis, but is not necessary for a journal article.
%
%The online Atomic Spectral Line Database\footnote{http://www.pmp.uni-hannover.de/cgi-bin/ssi/test/kurucz/sekur.html} was used to identify ions that could have caused certain absorption features. The relative cosmic abundance and %energy levels of the listed ions are used to determine which ion is the likely cause of %the absorption line. Debris disks have temperatures of order ten, so ions that have a %high baseline energy when transitioning are unlikely suspects.

In addition to the UV spectra, we obtained optical spectra of \cet\ using the Sandiford Echelle Spectrograph on the 2.1-meter telescope at the McDonald Observatory. 
Five observations were recorded: one on 2013-08-09, three on 2013-08-11 (the day of 
STIS Visit~1), and one on 2013-08-12.
The spectra covered the wavelength region around the \ion{Ca}{2} K line (3933~\AA) 
at a spectral resolution of $R \sim 60,000$.
These data are discussed in Section~\ref{sub:ca2}.

%%%%%%%%%%%%%%%%%%%%%%%%%%%%%%%%%%%%%%%%%%%%%%%%%%%%%%%%%%%%%%%%%%%%%%%%%%%%%%%%%%%%%%%%%%

\section{Analysis}

For an absorption feature caused by optically thin gas with complex velocity structure, the apparent optical depth method can be used to measure (or limit) the absorbing column densities in particular velocity ranges \citep[e.g.][]{Roberge:2002}.
Following \citet{Savage:1991}, the equation for the column density over a specific velocity range is 
\begin{equation}
N = \frac{m_{e} c}{\pi e^{2} \lambda_{0} f } \int_{v_{1}}^{v_{2}} \ln \frac{ I_{0}(v)}{ I_{m}(v) } \: dv \: ,
\end{equation}

where $m_{e}$ is the mass of an electron, $c$ is the speed of light, $e$ is the charge of an electron, $\lambda_{0}$ is the central wavelength of the absorption line, $f$ is the line oscillator strength, $I_{0}(v)$ is the intensity of the light from the star without superimposed absorption (i.e.\ the continuum), and $I_{m}(v)$ is the intensity measured after light from the star has traveled through some medium along the path to the observer (i.e.\ the measured spectrum).

Assuming the intensity of the star does not vary on timescales of days, an increase in the {\bf column density} along the line of sight will lead to a decrease in observed intensity at specific wavelengths. The change in column density can be measured from the ratio of the intensities measured during and before or after the event, integrated over the appropriate velocity range.

\begin{equation}
N_{2} - N_{1} = \frac{m_{e} c}{\pi e^{2} \lambda_{0} f }  \int_{v_{1}}^{v_{2}} \Big( ln \frac{ I_{0}(v)}{ I_{m,2}(v) } -  ln \frac{ I_{0}(v)}{ I_{m,1}(v) } \Big)   dv 
\end{equation}

\begin{equation} \label{eq:eq}
\Delta N = \frac{m_{e} c}{\pi e^{2} \lambda_{0} f } \int_{v_{1}}^{v_{2}} \ln \left( \frac{ I_{m,1}(v)}{ I_{m,2}(v) } \right) dv 
\end{equation}
The column density difference $\Delta N$ is the material associated with variable gas. This analysis was done for every variable feature, except the unidentified feature. 
The $1 \sigma$ uncertainties on the $\Delta N$ values were determined by propagating the flux measurement errors through Equation~\ref{eq:eq}.
% A variable feature is a change in flux greater than 3$\sigma$, over a large wavelength 
% range.

%%%%%%%%%%%%%%%%%%%%%%%%%%%%%%%%%%%%%%%%%%%%%%%%%%%%%%%%%%%%%%%%%%%%%%%%%%%%%%%%%%%%%%%%%%

\section{Variable Species}

\subsection{\ion{C}{2}}

\begin{figure}
\plotone{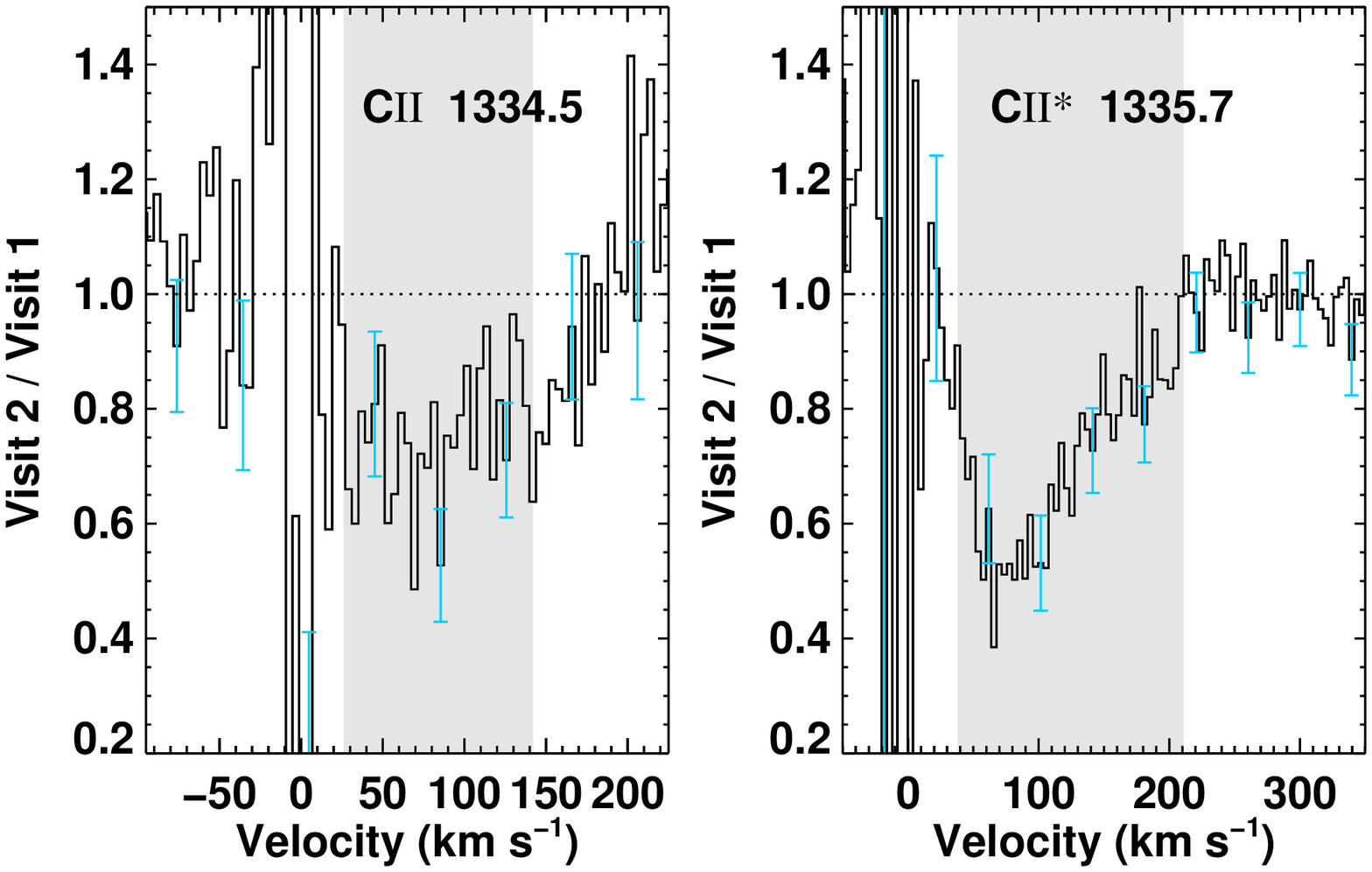}
\caption{Details of the variable \ion{C}{2} absorption events.
In the left panel, the Visit~2 spectrum divided by the Visit~1 spectrum is plotted, 
isolating the redshifted excess absorption visible on the wing of the 1334.5~\AA\ line.
The right panel also shows the Visit~2 spectrum divided by the Visit~1 spectrum, 
isolating the redshifted excess absorption visible on the wing of the 1335.7~\AA\ line.
The x-axis shows the velocity relative to that of the central star. 
The velocity ranges over which excess absorption is significantly detected (given in Table~\ref{tab:lines}) are indicated with gray bars.}
\label{fig:cii_zoom}
\end{figure}

In the Visit~2 spectrum, excess absorption is seen on one wing of both the \ion{C}{2} and \ion{C}{2}* lines. 
Figure~\ref{fig:cii_zoom} highlights this absorption with plots of the Visit~2 spectrum normalized by the Visit~1 spectrum. The excess absorption is redshifted with respect to the star, showing that the gas was infalling towards the star at the time of transit. The lower energy levels of the two lines are not greatly different from each other, and the shapes of the excess absorption in both lines are also similar. 
Therefore, the variable \ion{C}{2} and \ion{C}{2}* likely arise from the same parcel of infalling gas.
The distinctive triangular shape of the excess absorption is very similar to the shape of variable atomic absorption seen in UV spectra of $\beta$~Pic and attributed to transits of star-grazing planetesimals \citep[e.g.][]{Vidal-Madjar:1994,Roberge:2000}. 

The large variations in the intensity ratio near $v = 0$ relative to the star's velocity are due to the very small measured intensity values at the bottoms of the saturated absorption lines from the main disk gas (see Figure~\ref{fig:cii}). These portions of the data were excluded from our analysis.
The \ion{C}{2}* line at 1335.7077~\AA\ overlaps with a low energy absorption line due to \ion{Cl}{1} at 1335.7258 \AA. 
However, there is another \ion{Cl}{1} line at 1347.2396~\AA\ arising from the same energy level, which is weak and invariant. 
Therefore, \ion{C}{2}* is responsible for most of the variable absorption and its column density was calculated using only the parameters of the \ion{C}{2}* 1335.7077~\AA\ line. 
Table~\ref{tab:lines} lists all of the variable carbon features analyzed using the apparent optical depth method, the velocity ranges over which the variable absorption was significantly detected, and the total column densities in the variable gas. 

\begin{deluxetable*}{lccccccc}
\tablecaption{Variable features and atomic data \label{tab:lines}}
\tablecolumns{8}
\tablehead{\colhead{Species} & \colhead{$\lambda_{0}$ \tablenotemark{a}} & \colhead{$E_\mathrm{lower}$ \tablenotemark{b}} & \colhead{$f$ \tablenotemark{c}} & \colhead{Appearance \tablenotemark{d}} & \colhead{$\Delta v$ \tablenotemark{e}} & 
\colhead{$\Delta N$ \tablenotemark{f}} & \colhead{Max.\ Distance \tablenotemark{g}} \\
\colhead{ } & \colhead{(\AA)} & \colhead{(cm$^{-1}$)} & \colhead{ } & \colhead{ } & 
\colhead{(km s$^{-1}$)} & \colhead{($10^{14}$ atoms cm$^{-2}$)} & \colhead{(AU)}}
\startdata
\ion{C}{2}    & 1334.5323         & 0            & 0.128  & Visit 2          & 26 -- 143         & 1.145  $\pm$ 0.380    & 0.17   \\
\ion{C}{2}*   & 1335.7077         & 63.42        & 0.115 & Visit 2          & 38 -- 211          & 2.115  $\pm$ 0.508    & 0.08   \\
\ion{C}{4}    & 1548.204         & 0    		   & 0.190  & Visit 1          & $-372$ -- $-10$  & 1.420  $\pm$  0.154   &  \nodata   \\
\ion{C}{4}    & 1550.781          & 0            & 0.095 & Visit 2          & 13 -- 258         & 1.603  $\pm$  0.196    & 0.05  
\enddata
\tablenotetext{a}{Rest wavelength}
\tablenotetext{b}{Energy of lower level of transition}
\tablenotetext{c}{Oscillator strength \citep{Morton:2003}}
\tablenotetext{d}{Visit that shows excess absorption}
\tablenotetext{e}{Velocity range of excess absorption}
\tablenotetext{f}{Column density in variable gas}
\tablenotetext{g}{Maximum distance of gas from star at time of transit}
\end{deluxetable*}

\subsection{\ion{C}{4}}

\begin{figure}
\plotone{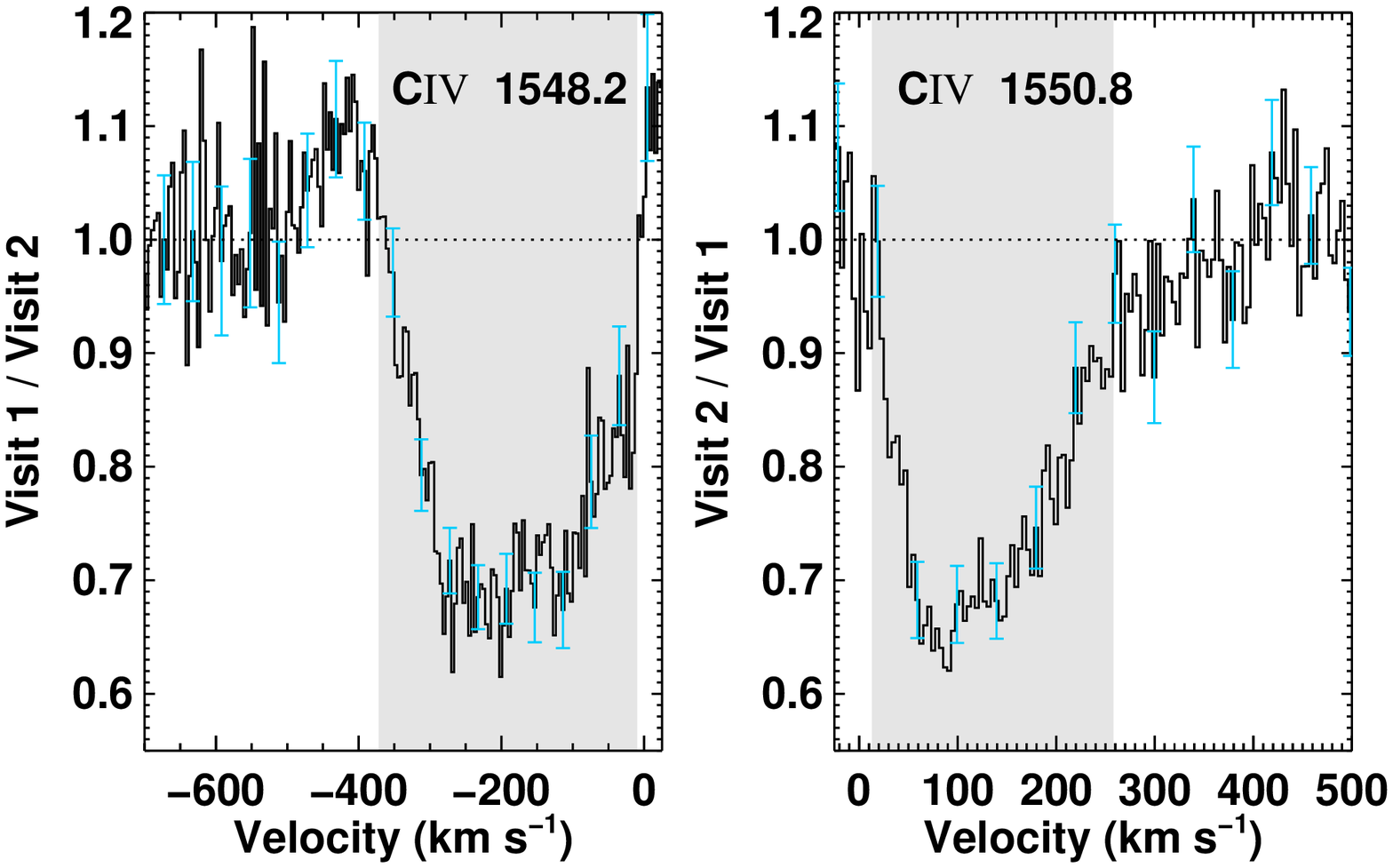}
\caption{Details of the variable \ion{C}{4} absorption events. 
In the left panel, the Visit~1 spectrum divided by the Visit~2 spectrum is plotted, 
isolating the blueshifted excess absorption visible on the wing of the 1548.2~\AA\ line.
The right panel shows the Visit~2 spectrum divided by the Visit~1 spectrum, 
isolating the redshifted excess absorption visible on the wing of the 1550.8~\AA\ line.
The x-axis shows the velocity relative to that of the central star.
The velocity ranges over which excess absorption is significantly detected (given in Table~\ref{tab:lines}) are indicated with gray bars. \label{fig:civ_zoom}}
\end{figure}

The two \ion{C}{4} lines show variable absorption on both the red and blueshifted wings, highlighted in Figure~\ref{fig:civ_zoom}. 
It appears there was an outgoing event in Visit~1 and an infalling one in Visit~2.
The presence of both red and blueshifted events is difficult to explain with other scenarios that do not involve star-grazing comets. 
For example, gas accretion should only produce redshifted absorption, while stellar winds or outflows should only produce blueshifted absorption. 
Furthermore, a species as highly ionized as \ion{C}{4} cannot be produced by photoionization in the CS environment of an A star.
As in the case of the highly ionized, variable species seen in \bp\ spectra, the \ion{C}{4} must be produced by collisional ionization in hot, dense gas.
Such conditions could occur in the shock at the leading edge of a star-grazing comet coma 
\citep{Beust:1993}.

The red and blueshifted \ion{C}{4} features are badly blended in the region between the two lines, making that portion of the data impossible to analyze. 
Fortunately, both lines arise from the same lower energy level ($0 \ \mathrm{cm}^{-1}$). 
Therefore, we could analyze the unblended blueshifted event in the \ion{C}{4} line at 1548.2~\AA\ and the unblended redshifted event in the \ion{C}{4} line at 1550.8~\AA\ to obtain clean measurements of $\Delta N$ for both events.
There is a hint of additional blueshifted excess absorption in the Visit~2 spectrum, 
apparent as a Visit~1 / Visit~2 ratio greater than 1 near $v \sim - 425$~\kms\ (see left 
panel of Figure~\ref{fig:civ_zoom}).
This may be a sign of an additional weak outgoing event in Visit~2 simultaneous 
with the strong infalling event. 
However, as the feature is only seen in one line and is not robustly detected, 
no definitive conclusions can be drawn.

How are the variable gas events seen in \ion{C}{2} and \ion{C}{4} related?
 In each visit, all spectra were obtained in two adjacent \emph{HST} orbits.
The first exposure containing the \ion{C}{2} lines was taken about 1.5~hours before the exposure containing the \ion{C}{4} lines. 
Therefore, the data from each visit are nearly simultaneous.
The similar shape and velocity range for the redshifted absorption seen in the Visit~2 data for both species make it likely that the features are associated with the same infalling gas event.
In contrast, the blueshifted absorption seen only in \ion{C}{4} during Visit~1 is detected over a wider velocity range, extending to much larger blueshifted velocities. 
Its higher velocity suggests the outgoing gas was closer to the central star at the time of transit than the infalling gas event.
This may explain why the blueshifted event was not seen in the \ion{C}{2} lines; the gas may have been hotter and more highly ionized overall. 
The different viewing geometry for the blueshifted event may also have contributed to its non-detection in \ion{C}{2}.

\subsection{Mystery Feature}

The mystery feature shown in Figure~\ref{fig:mystery} is unlike the other variable features, since it appears that there is no stable unvarying absorption at all.
This made it difficult to identify the species responsible for the absorption feature.
However, the general shape of the variations does resemble the variations seen in the \ion{C}{4} lines: redshifted excess absorption in Visit~2 and blueshifted in Visit~1.
This feature is unlikely to arise from \ion{Fe}{2}, as the 1608.46~\AA\ absorption line is seen in the dataset but does not show significant variation. In far-UV spectra of \bp\ showing variable features, highly ionized species like \ion{C}{4} generally appear more strongly variable than less ionized species like \ion{C}{2} \citep[e.g.][]{Bouret:2002}.
Therefore, the fact that all of the mystery absorption appears to be variable leads us to suspect that it is due to a very highly ionized and/or energetic ion.
%One candidate is \ion{Mo}{6} at 1595.435~\AA, but we are unable to confirm this
%identification.

\subsection{\ion{Ca}{2} \label{sub:ca2}}

\begin{figure}[b!]
\plotone{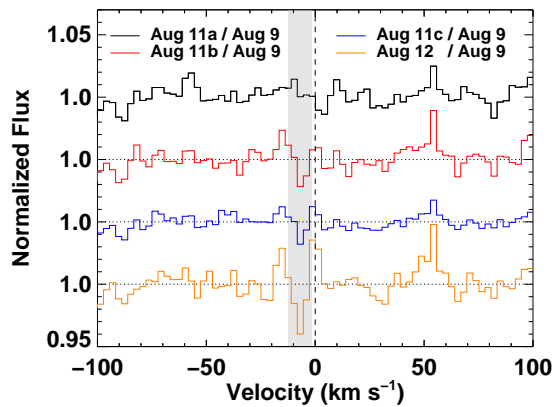}
\caption{The \ion{Ca}{2} K line in ground-based optical spectra of \cet.
The y-axis shows the normalized fluxes, obtained by dividing the spectra taken 
on 2013-08-11 (Aug~11a, Aug~11b, Aug~11c) and 2013-08-12 (Aug~12) by the spectrum 
taken on 2013-08-09 (Aug~9).
For clarity, each ratio spectrum was shifted by a constant normalized flux value 
before plotting.
The x-axis shows the velocity relative to that of the central star. 
A weak blueshifted absorption feature appears in the Aug~11b, Aug~11c, and Aug~12
spectra. The variable region is highlighted with the gray bar.
\label{fig:ca_2}}
\end{figure}

The optical spectra of the \ion{Ca}{2} K line (3933~\AA) are shown in 
Figure~\ref{fig:ca_2}.
To highlight absorption variations, the three spectra taken on Aug~11 (the day of
STIS Visit~1) and the spectrum taken on Aug~12 were divided by the spectrum
taken on Aug~9.
No significant variation was seen between the first spectrum taken on Aug~11 (Aug~11a) 
and the Aug~9 reference spectrum. 
However, a weak blueshifted absorption feature at $v \sim -8$~\kms\ appears in 
the Aug~11b, Aug~11c, and Aug~12 spectra.
At first blush, this feature could be associated with the blueshifted absorption 
feature in the \ion{C}{4} profiles from Visit~1 (see Figure~\ref{fig:civ}).
However, the large difference in velocity shift between the \ion{Ca}{2} and \ion{C}{4} variable features indicates that the low-ionization and high-ionization gas are 
not closely associated with each other. 
It seems likely that there is a velocity and ionization gradient within each variable gas event, or that each event may actually be produced by more than one star-grazing 
comet, something that has been revealed in ultra-high resolution \ion{Ca}{2} 
spectra of \bp\ \citep{Crawford:1994}.

%%%%%%%%%%%%%%%%%%%%%%%%%%%%%%%%%%%%%%%%%%%%%%%%%%%%%%%%%%%%%%%%%%%%%%%%%%%%%%%%%%%%%%%%%%

\section{Discussion}

\subsection{Velocity and Stellar Distance}

The velocity of the gas as it transits the host star can be used to find the 
maximum distance possible of the comet(s) from the star at a given time.
An object that is gravitationally bound to a star must travel at or below 
the free-fall velocity, given by 
\begin{equation}
v_{ff} = \sqrt{\frac{2 G M}{r}} \: ,
\end{equation}
where $r$ is the distance of the object, $G$ is the gravitational constant and $M$ is the mass of the central star. 
Other massive bodies like planets can speed up or slow down objects, but for this analysis we considered such perturbations to be negligible. 
All other forces on infalling (redshifted) gas, like radiation pressure, will act to reduce its radial velocity \citep{Beust:1989}.
Therefore, the free-fall velocity is the maximum velocity that gas coming from an orbiting body moving towards the central star can have.
Adopting a mass for this A1V star of $2.7~M_\sun$, the fastest velocities for all of the redshifted events correspond to maximum distances ranging from 0.05 to 0.17~AU, very close to the star. 
% Perihelion of Halley’s Comet occurs at 0.5 AU.

\subsection{Upper Limits on Abundances of Other Atoms}

None of the other CS absorption lines arising from the main disk gas significantly varied between the two visits.
We analyzed unvarying lines arising from several species (\ion{O}{1}, \ion{Al}{1}, \ion{Si}{2}, \ion{S}{1}, \ion{Cl}{1}, and \ion{Fe}{2}) in order to set upper limits on the amounts of these gases coming from star-grazing comets and compare them to the variable gas abundances.
The specific lines analyzed appear in Table~\ref{tab:abundances}.
In each case, we placed upper limits using the strongest line of the species appearing in the dataset.

%For this comparison analysis, we focused only on the redshifted carbon events, for which %we have the most information.
For this comparison analysis, we analyzed only the range of velocities over which redshifted (i.e.\ infalling) gas was detected in all the variable lines (38 -- 143~\kms).
This increases the likelihood that we are analyzing the same parcel of gas for all species.
We applied Equation~\ref{eq:eq} to all the lines, including the variable ones, and used the flux uncertainties to set $3 \sigma$ upper limits on the unvarying gas in this velocity range.
The results appear in Table~\ref{tab:abundances}.

To accurately measure total elemental abundances in the infalling gas, we would need to analyze lines from many ionization states of each element. This is not possible, since not all of the necessary lines appear in the data. A measurement of the temperature in the infalling gas would permit an estimate of the total elemental abundances assuming collisional ionization equilibrium.
Assuming local thermodynamic equilibrium, the ratio of the \ion{C}{2} and 
\ion{C}{2}* column densities in Table~\ref{tab:abundances} implies a temperature of
$\sim 1000$~K. However, this is a lower limit on the true temperature in the infalling gas, since there could be more \ion{C}{2} gas in lines at higher energy levels that we cannot observe. Furthermore, this temperature is obviously too low to produce \ion{C}{4} by collisional ionization. Nor can we accurately estimate the temperature in the infalling gas from the abundance ratio of \ion{C}{2} and \ion{C}{4}, since \ion{C}{2} may be produced by photoionization as well as collisional ionization.

However, it appears that \ion{C}{2} is a dominant ionization state of carbon in the infalling gas, since no variable \ion{C}{1} is seen (although strong CS \ion{C}{1} lines appear in the data) and the variable \ion{C}{4} is about a factor of three less abundant. 
In such gas, \ion{O}{1} is likely to be an abundant state of oxygen, since its first ionization potential (13.62~eV) is higher than that of carbon (11.26~eV).
Furthermore, C and O both feel weak radiation pressure from the central A star \citep{Fernandez:2006}. These elements should be similarly affected by dynamical forces and should not be spatially segregated.
Therefore, we conservatively chose to represent the total variable carbon abundance by the sum of the \ion{C}{2} $\Delta N$ values ($\Delta N_C = (2.65 \pm 0.47) \times 10^{14} \ \mathrm{cm}^{-2}$) and compared that to the upper limit on the \ion{O}{1} abundance. 

With these assumptions, the ratio of C to O in the infalling gas is $\gtrsim 1.5$, at least 3 times the solar value \citep[solar C/O = 0.5;][]{Lodders:2003}.
A carbon overabundance relative to O is also seen in the \cet\ stable disk gas \citep[C/O $\gtrsim 4.5$;][]{Roberge:2014}.
Since sub-mm CO emission is seen from \cet\ \citep{Hughes:2008}, one is tempted to consider whether CO-rich planetesimals could be responsible for the variable gas.
No CO absorption was seen in the STIS spectra of \cet\ \citep{Roberge:2014}, likely due to the fact that the disk is not exactly edge-on (Lieman-Sifry \& Hughes, in preparation).
Obviously, there is also no detectable CO in the variable gas.

Given the close distances to the star of the variable gas at the time of transit, any molecular species would likely be dissociated.
If CO was the sole source of the variable atomic gas, one would expect a C/O ratio close to 1, which does not appear consistent with the lower limit on the C/O ratio in the redshifted gas. However, given the uncertainties about the ionization balance in the infalling gas, this conclusion is highly tentative. We note that in the case of \bp, the measured line-of-sight abundances of C and CO indicate that CO cannot be the sole source of carbon in the bulk disk gas \citep{Roberge:2000}.

\begin{deluxetable}{lcccc}
\tablecaption{Abundances in the Visit~2 redshifted gas \label{tab:abundances}}
\tablecolumns{5}
\tablehead{\colhead{Species} & \colhead{$\lambda_{0}$ \tablenotemark{a}} & 
\colhead{$E_\mathrm{lower}$ \tablenotemark{b}} & \colhead{$f$ \tablenotemark{c}} & 
\colhead{$\Delta N$ \tablenotemark{d}} \\
\colhead{ } & \colhead{(\AA)} & \colhead{(cm$^{-1}$)} & \colhead{ } & \colhead{($10^{14}$ atoms cm$^{-2}$)}} 
\startdata

\ion{C}{2}    & 1334.5323         & 0            & 0.128  & 0.997 $\pm$ 0.324 \\
\ion{C}{2}*   & 1335.7077         & 63.42        & 0.115  & 1.654 $\pm$ 0.335 \\
\ion{C}{4}    & 1550.781          & 0            & 0.095  & 0.917 $\pm$ 0.008 \\
\ion{C}{1}**  & 1561.4384         & 43.50        & 0.0675 & $\leq$ 0.387 \\
\ion{O}{1}    & 1302.1685         & 0            & 0.049  & $\leq$ 2.332 \\
\ion{Al}{2}   & 1670.7874         & 0            & 1.833  & $\leq$ 0.017 \\
\ion{Si}{2}   & 1304.3702         & 0            & 0.147  & $\leq$ 2.200 \\
\ion{S}{1}    & 1425.0299         & 0            & 0.192  & $\leq$ 0.326 \\
\ion{Cl}{1}   & 1347.2396         & 0            & 0.119  & $\leq$ 0.480 \\
\ion{Fe}{2}   & 1608.4511         & 0            & 0.062  & $\leq$ 0.318 
\enddata
\tablenotetext{a}{Rest wavelength}
\tablenotetext{b}{Energy of lower level of transition}
\tablenotetext{c}{Oscillator strength \citep{Morton:2003}}
\tablenotetext{d}{Column density in variable gas over velocity range 38 -- 143~\kms}
\end{deluxetable}

\section{Conclusion}

Our observations strengthen the connection between the variable gas in the \cet\ disk and star-grazing comets.
First, both red and blueshifted events are seen, which is hard to explain with other scenarios.
Second, the distance of the gas at the time of transit is within about 0.2~AU of the central star.
The non-detection of variable \ion{O}{1} features suggests that the C/O ratio in the gas is super-solar, too high for CO to be the primary source of the gas. Detailed modeling of the ionization balance in the variable gas will be needed to confirm this suggestion and to determine the abundance ratios of other elements.
Comets and other small bodies play an active role in the development of planetary systems, as we have long known for our own Solar System. As a young system with an apparently large and active comet population, \cet\ is a vital debris disk for learning about planetesimals in the context of planetary formation and evolution. 

%From our observations, the star-grazing comets within the 49~Cet debris disk appear to be %very carbon rich. No Doppler shifted oxygen or CO appeared in the data and upper limits %on un-detectable oxygen were not significant to the amount of carbon actually detected in %the disk. If line of sight CO is not detectable, more spectroscopic observations need to %be made to determine if these comets are oxygen depleted or not. While we do not have a %direct link between the presence of CO and comets, we do have new information concerning %the composition of comets in another planetary system. Comets and other small bodies play %an active role in the the development of planetary systems, as we’ve seen in our own %Solar System. As a young system with a predicted large and active comet population, %49~Cet is a vital debris disk for learning about comets in the context of planetary %formation and evolution. 

%%%%%%%%%%%%%%%%%%%%%%%%%%%%%%%%%%%%%%%%%%%%%%%%%%%%%%%%%%%%%%%%%%%%%%%%%%%%%%%%%%%%%%%%%%

\acknowledgments

Support for program number GO-12901 was provided by NASA through a grant from the Space Telescope Science Institute, which is operated by the Association of Universities for Research in Astronomy, Inc., under NASA contract NAS5-26555. A.~R.\ also acknowledges support by the Goddard Center for Astrobiology, part of the NASA Astrobiology Institute.

{\it Facilities:} \facility{HST (STIS)}

%\bibliographystyle{apj}
%\bibliography{bibliography}

\end{document}